# Crossing the hurdle:

## the determinants of individual scientific performance

A. Baccini, L. Barabesi, M. Cioni, C. Pisani

Department of Economics and Statistics - University of Siena

An original cross-sectional dataset referring to a medium-sized Italian university is implemented in order to analyze the determinants of scientific research production at individual level. The dataset includes 942 permanent researchers of various scientific sectors for a three-year time-span (2008-2010). Three different indicators - based on the number of publications and/or citations - are considered as response variables. The corresponding distributions are highly skewed and display an excess of zero-valued observations. In this setting, the goodness-of-fit of several Poisson mixture regression models are explored by assuming an extensive set of explanatory variables. As to the personal observable characteristics of the researchers, the results emphasize the age effect and the gender productivity gap - as previously documented by existing studies. Analogously, the analysis confirm that productivity is strongly affected by the publication and citation practices adopted in different scientific disciplines. The empirical evidence on the connection between teaching and research activities suggests that no univocal substitution or complementarity thesis can be claimed: a major teaching load does not affect the odds to be a non-active researcher and does not significantly reduce the number of publications for active researchers. In addition, new evidence emerges on the effect of researchers administrative tasks - which seem to be negatively related with researcher's productivity - and on the composition of departments. Researchers' productivity is apparently enhanced by operating in department filled with more administrative and technical staff, and it is not significantly affected by the composition of the department in terms of senior/junior researchers.

## Acknowledgments

We would like to thank Sonia Boldrini from the Evaluation Committee of the University of Siena for her valuable assistance in data harvesting and database building. In addition, we would like to express our gratitude to two anonymous referees for their comments which have led to a truly improved version of the paper.

## 1. Introduction

During the past thirty years, there has been a growing interest in the role of academic research activity and in its contribution to economic growth and social development. One of the least studied and most puzzling feature of this debate is the question of individual scientific productivity. Researcher's activity is basically a multi-output activity, producing outcomes as research, teaching and others products (newspapers articles, medical protocols, etc.) with a relevant impact on society. This idea is so widespread that in the national research assessment exercises - as an example in the British Research Excellence Framework 2014 - information is collected on all these different activities in order to evaluate not only the quality of research produced by universities, but even their multifaceted societal impact. At the best of our knowledge, no study addresses the question of the determinants of researchers' overall production, even if many papers solely focus on one dimension of their multi-output activities.

The idea that scientific publications represent the essence of the research activity is widely accepted (Wooton 2013). In this respect, two different streams of literature have been developed. The first stream is focused on describing the laws underlying the frequency distribution of researcher's publications - following the tradition started by Lotka (1926). The second stream deals with the determinants of individual productivity. These works aim to define the factors affecting research productivity by using tools as correlation analysis or regression modelling.

A common feature of this literature consists in considering the number of publications of a researcher as a proxy for quantifying her/his productivity. The main drawback of this indicator stems on the fact that each publication counts for one: a short paper addressing a limited issue counts as a seminal paper. So, the holy grail of scientometric research is the construction of indicators addressing - at the same time - the issue of productivity and quality of scientific work also at a researcher level (van Leeuwen et al. 2003). To this aim, the most-used strategy in empirical research consists in the substitution of the notion of "research quality" with the notion of "scientific impact" (as defined in Martin and Irvine 1983) which can be more easily handled using citation data. A first possible approach may be based on counting a subset of the publications of a researcher, such as the highly-cited papers or those published in "top" journals - so defined in reference to the impact factor or to other similar indicators. A different approach aims to obtain composite indicators of productivity and impact of a researcher considering her/his published articles as well as the citations they received. Among them, the most widespread indicator is the h-index (Hirsch 2005).

The aim of this article is to contribute to the debate on the determinants of individual scientific production by stressing the attention on two original issues. The first issue deals with the adoption of regression models able to properly handle the specific nature of the data. Three standard different indicators, based on the number of publications and/or their citations, are considered as response variables. They are integer-valued and display highly-skewed distributions which are in addition zero-inflated, i.e. an excess of zeroes is present. Many existing papers dealing with modelling the determinants of scientific production do not specifically account the skewed and zero-inflated nature of the data (Carayol and Matt 2004, 2006; Lissoni et al. 2011; Rivera-Huerta et al. 2011). In order to address the skew distribution of research output - and eventually acknowledging the presence of zero excess - quantile regression may be adopted (Kelchtermans and Veugelers 2011). In the present paper, an alternative approach, based on appropriate GLM models expressly conceived for count data with zero excesses, is proposed. Quite surprisingly some of these models, even if well-established in other disciplines, have been neglected in the framework of the analysis of production process of academic research. The second issue focuses on the joint use of an extensive set of explanatory variables, which have been considered by adopting separate analyses in the previous literature. Indeed, the proposed models consider as possible determinants of the researcher's individual academic productivity: (i) some personal observable characteristics, such as gender and age; (ii) some individual career features, such as academic position, seniority, typology of labour contract; (iii) the scientific field in which the scholar works; (iv) her/his teaching and administrative tasks; (v) her/his departmental working context. These ideas are carried out on a large original dataset referred to a set of about one thousand scholars belonging to a medium-sized Italian university.

The remainder of the paper is subdivided into four sections. In Section 2, the main literature addressing the issues of researcher's production is surveyed. In Section 3, the data and the methodologies adopted are illustrated. In Section 4, the main results of our empirical analysis are presented and discussed. Finally, in Section 5 the conclusions are drawn.

## 2. Literature review

Two alternative approaches have been considered in the analysis of scientific production: the first focuses on the laws underlying the frequency distribution of the number of publications (or citations), while the second aims to identify the determinants of scientific performance. The first approach dates back to the 1920s, particularly to the publication of Lotka's seminal article (1926). Lotka investigated the frequency distribution of scientific productivity of chemists and physicists showing that "...the number (of authors) making $n$ contributions is about $1/n^2$ of those making one; and the proportion of all contributors, that make a single contribution, is about 60%...". This kind of approach has survived until the recent attempts to create theoretical models able to foresee the future pattern of the production of a scholar given her/his past performance (e.g. Wang et al. 2013 and the bibliography cited thereon).

Alternative explanations of the Lotka findings and - more generally - of the highly-skewed nature of scientific production of scholars have been proposed. The simplest one - highly criticized for example by Allison and Stewart (1974) and by David (1994) - is the so-called "sacred-spark hypothesis", i.e. the differences in productivity reflect unequal and predetermined capabilities of researchers. In the late 1960s, a so called "Matthew-effect hypothesis" was advanced by Merton (1968). Merton highlighted that well-known researchers receive more recognition for their work than less known researchers. Subsequently, this hypothesis was generalized by Cole and Cole (1973) to be valid not only for recognition, but even for scientific productivity. In this form, this hypothesis was called "cumulative advantage hypothesis". The idea is that recognition received early in researchers' career may be reinforced over time as it would enable easier access to research resources - this issue means that any advantage will be cumulative (Defazio et al. 2009). This kind of explanation exclusively focuses on the social structure in which scholars are embedded and work.

A second approach addresses the academic research production aiming to identify the determinants of scientific productivity. In these works the sacred-spark and the Mattew-effect hypotheses are considered as residual or unexplained components of a roughly defined production function addressing all the relevant explanatory variables. This explanatory approach has been applied both to the individual level of analysis - where the survey unit is the researcher - and to the aggregate level - where the survey unit is the research unit - e.g. department, laboratory, university. A complete review of the elements that a broad and growing applied literature considers as possible explanatory variables of individual and aggregate scientific productivity is not the aim of our work. Nevertheless, it can be suggested a possible twofold categorization, which separates individual determinants, referring to each scholar's characteristics, and collective determinants, relating to the features of the organization in which she/he is working.

A first group of individual determinants properly refers to the personal characteristics of a researcher, such as gender and age. Regarding the role of gender in scientific publication performance, at least starting from Cole and Zuckerman (1984), the gender differences in productivity among academic scientists is considered as a puzzle to be solved (Levin and Stephan 1998; Xie and Shauman 2003; Fox 2005; Leahey 2006; Fox et al. 2011). Mairesse and Pezzoni (2013) revisit the gender gap in scientific production, offering a critical review of the empirical evidence throughout the analysis of the issues influencing women scientific productivity (family engagements, marital status and policies in favour of women, institutional specificities, discipline specificities, etc.). Abramo et al. (2009) document differences in productivity between men and women, but highlight a progressive reduction of the performance gap over time for Italian scientists at least in hard sciences and life sciences. Similar conclusions are achieved by van Arensbergen et al. (2012) who suggest that - even if men outperform women in terms of scientific production in the older generations - the gendered differences are disappearing in the younger generations.

The so called age-effect is considered a well-consolidated issue in literature. Many studies document a decrease of research production as age increases (Diamond 1984, 1986); others find that publication activity tends to augment in the early career, reaches a peak, and then decreases (Zuckerman and Merton 1972; Weiss and Lillard 1982; Levin and Stephan 1991); while others find a productivity curve with two peaks (Bayer and Dutton 1977). These relationships have to be taken with caution, since it is difficult to distinguish between age effect and cohort effect. Indeed, the latter can be associated - for example - with a progression of knowledge or a different availability of resources, as discussed by Stephan (1996, 2012). At the aggregate level of analysis, a study on the Italian National Research Council highlights a negative relationship between age and research productivity indicators (Bonaccorsi and Daraio 2003), while the results by Carayol and Matt (2004) suggest an "inversed-U shape" relationship between laboratories productivity and age. In contrast, a lack of significant relationship between age and publication rate within the faculties of the University of Vienna is claimed by Wallner et al. (2003).

A second group of individual determinants refer to the career features. As to the effects of seniority and career progress, the role of tenure and position are ambiguous. On one hand, an improvement in the professional status of the researcher can positively affect research performance, since - for example - she/he can have easier access to funding or attract talented young students/scientists in her/his research team (the so-called "status effect"). On the other hand, once a career progress has been obtained, the incentives to production can be reduced. The relationship between seniority or career progress and research performance is nearly universally addressed in this literature. Among others, Fabel et al. (2008) find a negative effect of career age on publications for full professors, while Rivera-Huerta et al. (2011) consider career years as a control variable in modelling individual research output.

The scientific field in which the scholar works is considered as a determinant of productivity since it is well known that publishing activities and citation patterns vary among scientific disciplines. These differences - as emphasized by Anania and Caruso (2013) - are particularly relevant in many areas of Social Sciences and Humanities, where scientific productivity and citation practices typically yield fewer citations per paper. Two strategies are used for tackling this problem. The first strategy tends to limit the analysis on researchers working in homogeneous scientific fields (see e.g. Lissoni et al. 2011; Pezzoni at al. 2012). The second strategy aims to model the research performance of scientists from heterogeneous areas including control variables for the researcher scientific discipline - defined according to some available classification (see e.g. Carayol and Matt 2006; Rivera-Huerta et al. 2011).

The researcher's teaching activities and administrative tasks are also considered determinants of productivity. The main question addressed by literature is if the engagement in these activities may crowd out research. According to some authors (Fox 1992; Taylor et al. 2006), these activities conflict since the more productive researchers spend less time for teaching and students in general. A substitutive relationship of this kind is also documented by a paper on French professors in Economics (Kossi et al. 2013). Contrasting results are conveyed by Fabel et. al (2008) showing that higher teaching loads in terms of class sizes do not deteriorate research productivity of business economists in Germany and Switzerland, and by Kelchtermans and Veugelers (2011) highlighting that alternative activities have very small and mostly insignificant effects on research output for scientists employed at the KU Leuven. However, as suggested by Stephan (1996), given the collaborative nature of science, individual determinants solely represent a part of the drivers of scientific production. Thus, determinants relating to the organization in which the researcher operates have to be considered. Allison and Long (1990) highlight the role of prestigious departments in encouraging individual scientific productivity. The composition of laboratories or departments in terms of type of researchers (full professors, assistant professors, PhD students, etc.) and their average age are also considered in the literature (Carayol and Matt 2006), as well as the quality of colleagues' production (Mairesse and Turner 2006; Lissoni et al. 2011;), and the fundraising ability (Carayol and Matt 2006). At the aggregate level of analysis, some scholars concentrate on the organization size, showing a positive size effect on laboratory productivity (Cainelli et al. 2006; Fabel et al. 2008); others find that small-sized departments are more productive (Carayol and Matt 2004, 2006) while others focus on the effects of the composition and average age of the research unit (Bonaccorsi and Daraio 2003). On one side, senior researchers may enhance the productivity of the younger due to co-author works or informal contacts. On the other side, the younger can act as incentives to stimulate the research activities of the older (Carayol and Matt 2004, 2006).

## 3. Material and methods

### 3.1. Data

In order to address the issues raised in the previous section, our analysis is based on a large original dataset implemented by using either internal administrative sources or external sources. The data refer to the University of Siena. Established in 1240, it is one of the oldest publicly-funded universities in Italy. It is a medium-sized university with about 17,000 students covering 8 scientific areas: Arts and Humanities, Economics, Engineering, Law, Mathematical, Physical and Natural Sciences, Medicine and Surgery, Pharmacy, Political Sciences. The dataset is composed by 942 individual records referred to permanent researchers of the University of Siena. For each researcher, we collected information dealing with personal characteristics, research activity, teaching activity, administrative tasks and departmental affiliation. The data concerning personal characteristics and departmental affiliation refer to December 31st 2010. The data regarding research and teaching activities, as well as administrative tasks, refer to the three-year period 2008-2010. The choice of a three year time-span is due to the costly nature of manual collection of disperse administrative data, and to the growing difficulties of finding comparative and complete information in administrative files when further years are considered. A three-year period is a time-span sufficient to capture the normal activity of scholars. It permits to avoid yearly anomalies that could have arisen in reference to researchers' activities such as organizational breaks affecting the number of students or of thesis supervised. A three-year period also permits to avoid accidental conditions such as sabbatical or health leave, maternity, etc., which could have arisen by examining a single year. Finally, it also permits to avoid accidental zeroes for the response variables - arising by delays in data recording in the bibliographic archives considered or by a long time-lapse from submission to publication in peer-reviewed journals.

### 3.1.1 Response variables

In order to quantify publication production, three different response variables are considered:

i. number of publications in Anagrafe.UNISI, denoted by *repository*. The Anagrafe.UNISI is the institutional research repository of the University of Siena. It is used by researchers in order to record all their research outputs. It is therefore filled with a broad range of outputs from scientific publications to teaching materials and informative articles - such as newspapers and magazine articles. It is worthwhile to note that not all the research outputs recorded in repository are peer-reviewed, and that the researchers are responsible for registering and classifying their outputs. The response variable *repository* is constructed by counting - for each scholar - the number of authored or co-authored scientific outputs, recorded in the repository during the period 2008-2010 and classified as articles, books, chapters in books and conference proceedings;

ii. number of publications in Scopus, the bibliographic database developed by Elsevier (http://www.scopus.com), denoted by *scopus*. This response variable is the number of authored or co-authored publications registered in the Scopus database in the period 2008-2010. In order to get this information for each researcher, the last and first names and the affiliation were queried in the Scopus Author search. If the author's name was not unique, the results were refined to ensure that the correct publications were attributed to the researcher checking for the curriculum vitae and the list of publications available on the researcher personal web site. The database was accessed in July 2011;

iii. h-index score, described by the variable *h_index*. This index - introduced by Hirsh (2005) - combines the author's article count and citation count into a single value. According to Hirsch (2005 p.16572), the h-index "... gives an estimate of the importance, significance, and broad impact of a scientist's cumulative research contributions ...". The value of the h-index was extracted from the Scopus database and refers to December 31$^{st}$, 2010. The database was accessed in July 2011 using the same procedure adopted for the variable *scopus*.

The use of the two response variables *repository* and *scopus* is motivated by the recognized differences in publication patterns among different disciplines. In many disciplines such as Italian Literature or Law, research outputs traditionally consist of books, chapter in books and articles in national language. Scopus database does not cover at all, or covers very partially, these research outputs. As a consequence, data on research activities of social scientists and art-and-humanities scholars tend to be systematically undervalued by the variable *scopus*. Therefore, the institutional repository - containing a wider set of research outputs - allows to quantify the research production of these scholars more properly. Indeed, the response variable *repository* also considers data on the so-called national literature (Hicks, 2004).

Finally, the use of the three different measures of research outcome is supported by the values of the correlation coefficients computed on the data (0.53, 0.42 and 0.71 respectively for *repository* and *scopus*, *repository* and *h_index* and *scopus* and *h_index*).

### 3.1.2 Explanatory variables

The explanatory variables are collected from internal administrative files. Among the variables describing researcher personal characteristics, the variable *gender* is the first considered. The database also includes the variable *age*, i.e. the age of the researcher at December 31$^{st}$, 2010. The variable *tenure* is the number of years since a researcher got a permanent position in the Italian university system. The variable *position* gives the professional role in the University of Siena at December 31$^{st}$, 2010. Three different permanent positions are defined in the Italian university systems: "ricercatore universitario" (assistant professor), "professore associato" (associate professor) and "professore ordinario" (full professor). The recruitment and career promotion system, from assistant professor to associate or full professor, and from associate to full professor, are ruled by national laws and based - at least on principle - on scientific productivity. Research and teaching activities, as well as wages, are centrally defined by national legislation. Researchers do not bargain for wages and academic duties on an individual level. Indeed, the publicly-funded Italian university system is centralized and researchers of all ranks are considered civil servants, employed by the government through a selection made by commissions of senior peers (for a more detailed analysis on this issue, see Lissoni et al. 2011, 257-263 and Pezzoni et al. 2012, 708-209).

Italian researchers can choose between full-time or part-time academic positions. Part-time researchers have a teaching load of about one-half with respect to full-time researchers. It is mandatory that researchers involved in private practice, such as, for example, lawyers, engineers or architects, have a part-time contract. The wage for a part-time academic contract is about one-half of a full-time contract. The variable *full-time* differentiates researchers with full-time and part-time contracts at December 31st, 2010.

It must be pointed out that researchers in our database represent a single population, as they are affected by the same rules for hiring, career and academic duties, regardless of the discipline to which they belong to. Therefore, the adoption of separate regression analysis for the different disciplines in order to model their different publishing activities and citation practices, seems a forced strategy. Thus, we prefer to include the discipline in the model as a factor. To this purpose, the variable *erc* is constructed by manually reclassifying the scientific disciplines, as defined in the Italian academic system, into the European Research Council sectors: Life Sciences (LS), Physical and Engineering Sciences (PE), and Social Sciences and Humanities (SH).

Prin projects - acronym of "Progetti di Ricerca di Interesse Nazionale" - are three-year research projects granted by the Italian Ministry of University and Research and represent the unique universal research funding system for basic research in Italy. They are assigned on a competitive basis and open to all disciplines. Each project is evaluated by peer reviewers. Projects receiving positive evaluation are grouped according to the scientific fields and ranked on the basis of the evaluation given by peers. Given the shortage of funding, a small minority of the positively evaluated projects are effectively funded. The dummy variable *prin* indicates if a researcher has been involved at least in one positive-evaluated Prin project during the period 2008-2010. Therefore, it can be considered a proxy of the engagement of a researcher in active searching for funding, and of the ability to write well-evaluated project proposal. Hence, it is not a proxy of the financial resources available for scholars' research, since a positive evaluated project is not necessarily funded.

In order to tackle the question if teaching and research are complementary or conflicting activities, we gathered information on researchers teaching activity through three variables. The variable *teaching* is the average number of teaching hours per month during the period 2008-2010, computed excluding months on leave. The variable *thesis* is the number of bachelor and master dissertations supervised by each researcher in the period 2008-2010, while the variable *students* is the average number of students attending lectures in the same period. Similarly, we would assess if the time devoted to the governance could badly affect the individual productivity. To this purpose, we introduce the variable *presence_faculty_meeting*, i.e. the proportion of faculty meetings attended by a researcher during the period 2008-2010. As the number of faculty meetings is the same for each scholar belonging to the same Faculty, this variable can be considered as a rough proxy of the diligence of a researcher to face institutional duties.

Finally, we wonder if research productivity is affected by the research context and, particularly, by the characteristics of the departmental staff, i.e. the so-called "departmental effect". In order to contemplate the department composition, we include two variables, denoted by *more_junior_ratio* and *taw_ratio*. The former variable is researcher-specific and represents the ratio between the number of researchers in more junior ranks and the number of researcher in the same or more senior ranks. Thus, for all the assistant professors in a given department, the variable is the ratio between PhD students and research fellows and the overall number of assistant professors, associate professors and full professors, while for the associate professors the variable is the ratio between PhD students, research fellows and assistant professors, and the overall number of associate professors and full professors. Finally, for full professors, the variable is the ratio between PhD students, research fellows, assistant and associate professors, and the number of full professors. The *taw_ratio* variable is department-specific and represents the ratio of the number of non-research staff units to the number of permanent researchers. The non-research staff of a department includes technical and administrative workers. It is worth noting that the experimental science departments, where the staff is involved in the laboratory activities and in the management of well-funded research projects, present the highest values of this ratio.

Table 1 reports the definition of the explanatory and response variables adopted in the analysis, as well as some corresponding descriptive statistics, while Table 2 reports the correlation matrix of the quantitative explanatory variables.

TABLE 1 ABOUT HERE

TABLE 2 ABOUT HERE

*3.2 Method*

From Table 1, it is at once apparent that the distributions of the three response variables show the presence of a remarkable number of zeroes and a high level of skewness. When dealing with modelling skewed count data with an excess of zeroes, it is well known that the usual Poisson (P) regression can be inappropriate (see e.g. Schubert and Telcs 1989). Indeed, the data tend to exhibit over-dispersion, i.e. a larger variance than that predicted by the mean and a large number of zero counts. Therefore, Poisson regression can be considered as a benchmark. In order to address over-dispersion, the Negative Binomial (NB) regression can be alternatively used (see e.g. Rao 1980). However, when the major source of over-dispersion is related to an excess of zero counts, more flexible count data models - such as zero-inflated, hurdle models or more general mixture models - have to be adopted.

Actually, in what follows, several Poisson mixture models are considered, starting from zero-inflated and hurdle Poisson and Negative Binomial models. Since the Negative Binomial can be expressed as a Poisson mixture model - where the mixturing distribution is a Gamma law - these models have indeed a common base. Subsequently, we also consider two general Poisson mixture models by adopting the Sichel and Waring laws. Hence, the common rationale underlying our approach stems on the use of the Poisson law as the primary distribution. For a detailed discussion on this topic, see the classical monograph by Johnson et al. (2005: 351-373). For a recent discussion on more advanced Poisson mixture and compound Poisson models, see e.g. Barabesi and Pratelli (2014) and Marcheselli et al. (2008).

*3.2.1 Zero-inflated models*

Let us assume that $Y$ be the random variable representing the response variable and that $Y_1, \ldots, Y_n$ be a sample of $n$ stochastically independent counts. Under the zero-inflated models, the response variable is modelled as a mixture of a Dirac mass at zero and an integer-valued distribution - usually referred to as the count component. Thus, if $\theta$ represents a unknown parameter vector, the response variable $Y$ has an integer-valued distribution $f(k; z, \theta)$, with probability $1 - p(x)$, where $z$ and $x$ denote suitable covariate vectors), which is inflated by zeroes with probability $p(x)$. More precisely, if $z_i$ and $x_i$ denote the value of the covariate vectors for the *i*-th individual, the probability function of the random variable $Y_i$ is given by

$$P(Y_i = 0 | x_i, z_i) = p(x_i) + (1 - p(x_i)) f(0; z_i, \theta)$$

and

$$P(Y_i = k | x_i, z_i) = (1 - p(x_i)) f(k; z_i, \theta), \quad k = 1, 2, \ldots .$$

In the following, two regression models are actually considered: a logistic regression managing "inflated" zero counts and a log-linear regression managing the remaining zero and non-zero counts, i.e.

$$\log \frac{p(x_i)}{1 - p(x_i)} = x_i^T \beta$$

and

$$\log \mathrm{E}(Y_i | z_i) = z_i^T \gamma,$$

where $\beta$ and $\gamma$ denote parameter vectors to be estimated. Among zero-inflated models, the most widely applied one is arguably the Zero-Inflated Poisson (ZIP) model (see e.g. Lambert 1992; Bohning et al. 1999; Hall 2000; Dalrymple et al. 2003; Rathbun and Fei 2006) where the count component is assumed to display a Poisson distribution. However, count data may exhibit a high variability precluding the use of a Poisson distribution. In such a case, a Negative Binomial distribution can be assumed to describe the count component of the model, giving rise to the Zero-Inflated Negative Binomial (ZINB) model (see e.g. Rose et al. 2006; Minami et al. 2007; Zhang et al. 2012).

### 3.2.2 Hurdle models

The hurdle models, originally introduced by Mullahy (1986), are two-component models: the first component is constituted by a Dirac distribution at zero, while the second component - i.e. the count component - is a truncated integer-valued distribution modelling strictly positive counts. Thus, the probability function of the random variable $Y_i$ is given by

$$P(Y_i = 0 | x_i) = p(x_i)$$

and

$$P(Y_i = k | x_i, z_i) = (1 - p(x_i)) \frac{f(k; z_i, \theta)}{1 - f(0; z_i, \theta)}, \quad k = 1, 2, \ldots .$$

Similarly to the framework of zero-inflated models, $p(x_i)$ and $\mathrm{E}(Y_i|z_i)$ are generally modelled by means of the logit and log-linear regression, respectively. In this setting, the Hurdle Poisson (HP) model postulates that the count component has a truncated Poisson distribution. Alternatively, when dealing with a marked data variability, the count component can be modelled by means of a truncated Negative Binomial distribution giving rise to the Hurdle Negative Binomial (HNB) model (Dalrymple at al. 2003; Zhang et al. 2012).

It is worth noting that - even if the hurdle model may apparently resemble the zero-inflated model, since they are essentially a mixture of a Dirac mass at zero with a count distribution - their interpretation is rather different. Indeed, hurdle models assume that zero counts can solely arise with probability $p(x_i)$, while under zero-inflated models $p(x_i)$ represents the probability of getting "excess zeroes". More precisely, in the last case, zero counts may be obtained from the Dirac distribution as well as from the count component.

### 3.2.3 Mixture models

Loosely speaking, mixture models arise when considering a probability distribution whose parameters are in turn allowed to vary according to a further distribution, the so-called mixing distribution. More precisely, if $f_\lambda(k;z_i)$ denotes the probability function characterized by the parameter $\lambda$ corresponding to the primary distribution and $g(\lambda;\theta)$ denotes the probability density function corresponding to the mixing distribution depending on the vector of parameters $\theta$, the mixture probability function of the random variable $Y_i$ is given by

$$P(Y_i = k|z_i) = \int f_\lambda(k;z_i)g(\lambda;\theta)d\lambda, \quad k = 0,1,\ldots.$$

In such a case, over-dispersion may be handled by adopting a specific model. This issue leads to the theory of mixtures of Poisson distributions (Johnson et al. 2005). Indeed, the Poisson distribution is assumed to be the primary distribution - owing to its simplicity and intuitive appeal - while the mixing distribution is selected in order to be flexible enough for describing the main features of the data (Burrell and Fenton 1993). Among these models, the Generalized Waring Regression (GWR) model (see e.g. Irwin 1968; Schubert and Glänzel 1984, Burrell 2005 and the extended methodology proposed by Rodríguez-Avi et al. 2009) is obtained when the gamma product-ratio distribution is adopted as mixing distribution (Sibuya 1979). Under this model, the probability function of the random variable $Y_i$ is given by

$$P(Y_i = k | z_i) = \frac{\Gamma(a_i + \rho)\Gamma(h + \rho)}{\Gamma(\rho)\Gamma(a_i)\Gamma(h)} \frac{\Gamma(a_i + k)\Gamma(h + k)}{\Gamma(a_i + h + \rho + k)k!}, \quad k = 0,1,\ldots$$

where $h$ and $\rho$ are unknown parameters, with $h > 0$ and $0 < \rho < 1$, while $a_i = (\rho - 1)\mathrm{E}(Y_i | z_i)/h$ and - similarly to the zero-inflated and hurdle models - the log-linear function $\log \mathrm{E}(Y_i | z_i) = z_i^T \gamma$ is adopted.

The Generalized Inverse Gaussian Poisson distribution - also known as the Sichel distribution (Sichel 1985) - is obtained by adopting the Generalized Inverse Gaussian as mixing distribution. As pointed out by Burrell and Fenton (1993), it constitutes a very flexible - yet still manageable - model for describing count data with long tails. Rigby et al. (2008) parameterize the Sichel distribution in such a way that it can be easily interpreted and used for regression models, giving rise to the Sichel (S) model. Following Rigby et al. (2008), and modelling the mean as a function of the covariate vector, the probability function of the random variable $Y_i$ turns out to be

$$P(Y_i = k | z_i) = \frac{K_{\upsilon + k}(\alpha_i)(\mathrm{E}(Y_i | z_i)/c)^k}{K_\upsilon(1/\sigma)(\alpha_i \sigma)^{\upsilon + k}k!}, \quad k = 0,1,\ldots$$

where $\sigma > 0$ and $-\infty < \upsilon < +\infty$ are unknown parameters, while $K_\eta(\cdot)$ denotes the modified Bessel function of the third kind of order $\eta$. In addition, we assume that $c = K_{\upsilon + 1}(1/\sigma)/K_\upsilon(1/\sigma)$ and $\alpha_i = (1/\sigma^2 + 2\mathrm{E}(Y_i | z_i)/(c\sigma))^{1/2}$. In turn, the log-linear function has been considered in order to link $\mathrm{E}(Y_i | z_i)$ to the covariate vector $z_i$.

### 3.2.4 Parameters estimation

In the present study, the P, NB, ZIP, ZINB, HP, HNB, GWR and S models were considered. The vectors of parameters $\gamma$ and/or $\beta$ - and eventually the shape and/or scale parameters - were estimated by means of the maximum-likelihood method. The computational procedures needed for the estimation were carried out by means of the R software (R Development Core Team, 2012). The glm( ) function (Chambers and Hastie 1992) in the stats package and the glm.nb( ) function in the MASS package (Venables and Ripley 2002) were adopted with Poisson and Negative Binomial regression respectively. The presence of over-dispersion in the Poisson regression models fitted for the three response variables - i.e. *repository*, *scopus* and *h_index* - was confirmed by the results of the test performed using the function dispersiontest( ) implemented in the AER package. (p-value <0.001 for *repository*, p-value = 0.002 for *scopus* and p-value = <0.001 for *h_index*). The functions `zeroinfl( )` and `hurdle( )` in the `pscl` package (Zeileis et al. 2008) were used for dealing with zero-inflated and hurdle regression models. We estimated the parameters of the zero-inflated and hurdle models for the three response variables by adopting the same explanatory covariate vectors for the two components of the models. The function `gamlss( )` of the `gamlss` package was used for estimating the Sichel regression model with the parameterization proposed by Rigby et al. (2008), while the function `GWRM.fit()` of the `GWRM` package was adopted with the Generalized Waring Regression model.

### 3.3 Model selection

The P, NB, ZIP, ZINB, HP, HNB, GWR and S models were compared on the basis of the log-likelihood values, as well as on the Akaike Information Criterion (AIC). The log-likelihood and AIC values of the estimated models are reported in Table 3. The analysis of this table highlights that:

- zero-inflated and hurdle models produce very similar log-likelihood and AIC values;
- Negative-Binomial-based models exhibit a marked better performance than Poisson-based models for fitting purposes. Indeed, the simple Poisson model cannot account for the large proportion of zero counts, and, even if the zero-inflated and hurdle Poisson models can address this lack of fitting, they are not able to predict the nonzero frequencies correctly;
- the GWR model shows a good performance in terms of AIC values, owing to the reduced number of parameters; indeed in the GWR model the probability of zero counts is not modelled.

Particularly, when the response variable *repository* is considered, the largest log-likelihood value is achieved under the HNB model, which also gives rise to the smallest AIC value. When the response *scopus* is considered, the HNB model accomplishes the largest log-likelihood value even if the minimum AIC is associated with the GWR model. Similarly, GWR model gives rise to the lowest AIC value with the response variable *h_index*, even if the lowest value of the log-likelihood is reached by the ZINB model.

TABLE 3 ABOUT HERE

In Table 4 the observed frequencies are compared with the expected frequencies obtained under the ZINB, HNB and GWR models, which give rise to the best fitting.

TABLE 4 ABOUT HERE

Despite the good performance in terms of AIC values, it is at once apparent that under the GRW model the expected frequencies are rather far from the observed ones for the zero and the smaller variable values, even if the fitting improves in the right tail of the distribution. In contrast, the ZINB and HNB models give rise to expected frequencies of the zero counts which are respectively very close and identical to the observed ones, and provides expected frequencies rather similar to the observed ones for the smallest values. Indeed, the performance of the zero-inflated and hurdle models is nearly indistinguishable on the basis of the goodness-of-fit statistics and - as is common with zero-inflated skewed data (see e.g. Rose et al. 2006; Zhang et al. 2012) - also similar parameter estimates occur (the results are available from the authors on request). However, one model type may be more appropriate in order to describe the underlying generating data process. In our framework, zero-inflated models allow for zeroes to arise either from potentially productive or unproductive populations: unproductive researchers can never produce a research outcome, thus giving rise to structural zeroes, while those potentially productive can either produce or not produce a research outcome. In contrast, under hurdle models, all the researchers are considered potentially productive so that no structural zeroes are assumed, but solely some researchers *pass the hurdle* by authoring or co-authoring at least a research outcome in the three-year time-span considered. If a researcher has passed the hurdle, she/he becomes an *active* researcher, and her/his performance is described by the count component of the hurdle model. Since the Italian recruitment system is based on scientific productivity, as previously remarked, there is no reason to assume the existence of an unproductive population of researchers. Moreover, a three-year time-span is apt to reduce at a minimum the probability of occurrence of accidental zeroes due to anomalies. Hence - in presence of similar goodness-of-fit statistics - the use of hurdle models might be preferred. Figures 1, 2 and 3 report the observed distribution of the three response variables along with the predicted distribution obtained using the HNB model.

FIGURES 1, 2 and 3 ABOUT HERE

## 4. Results and discussion

In Table 5 we present the parameter estimates, their standard errors and the corresponding significance values when the HNB model is adopted (the same quantities for the other models are available from the authors upon request).

TABLE 5 ABOUT HERE

All the three measures of research output are significantly and negatively affected by *age* for the active researchers, i.e. those passing the hurdle. The relative decrease is approximately equal to 2% for all the three measures. Moreover, the variable *age* has a significant impact on the probability of having a zero h-index value, with a relative increase of the odds equal to 11%, as well as on the probability of having no publications recorded in the Scopus database, with a relative increase of the odds equal to 8%. These results are not in contrast with the prevailing evidence reported in the previous literature. Considering the cumulative nature of the h-index as a production measure, this finding is slightly puzzling. The issue might be interpreted as the result of two connected processes. The first refers to the coverage of the Scopus database. Older researchers - particularly the oldest - have publications dating back up to 35-45 years and which could have been appeared in journals not indexed in the Scopus database. Moreover, citations to oldest articles could not be recorded in Scopus. The second process refers to changes in publication strategies and citation habits over the years. Indeed, younger researchers may be more sensible to the publish-or-perish pressure, and they usually pay more attention to the outlet in which their works are published. In such a case, Impact Factor or other similar journal indicators drive the choice of target journals for submitting articles. Therefore, journals indexed in bibliographic databases, such as Scopus, are usually preferred with respect to non-indexed journals. Moreover, it is worth noting that, since science has become more and more connected in the recent years, younger scholars may have been able to achieve an higher h-index faster than in the past. A complete appreciation of these issues require longitudinal information not available in our database.

The gender effect is moderately significant and affects all the research production measures negatively. In fact, *gender* is slightly significant in the case of *repository* and *scopus* with respect to the count component of the model, and for the variable *h_index* when considering the zero component. This add another piece of evidence to the gender productivity gap, suggesting that women face *ceteris paribus* more difficulties than men in publishing. In turn, the not-longitudinal nature of our data does not permit to explore further this topic as done for example by Mairesse and Pezzoni (2013).

As to the academic position, its effect is significant for all the response variables. In particular, referring to the model count component, the number of publications and h-index value of the active researchers decrease for associate professors and assistant professors with respect to full professors (relative decrease respectively equal to 23% and 48%). Moreover, as evidenced by the zero component of the model, the academic position has a marked effect on the probability to be a non-active researcher. Indeed, when the variable *h_index* is considered, the increases of the odds of having a zero value for associate professors and assistant professors are approximately equal to 4% and 20% with respect to full professors. In turn, similar results hold for the variable *scopus*, while for the variable *repository* the effect is weakly significant on the odds only for assistant professors. The effect of *position* on research productivity can be related to the still-surviving hierarchical organization of the university which allows for full professors to act as both coordinators of national and international projects - which generally give rise to many co-authored publications - and supervisors for PhD students and junior researchers - who can stimulate their production.  It should be pointed out that, especially when the *h*-index value is considered, there might be a reverse explicative process between academic position and research output: an higher academic position should be determined by the life-long scientific production activity. This interpretation is coherent with the organizational characteristics of the Italian university system, in which promotions are mainly based on research activities. Moreover, this interpretation is also supported by the results by Lissoni et al. (2011) who proved that in Italy promotion is affected by the quantity of past publications.

As expected, the scientific sector of activity significantly impacts on research performance: the researchers who have passed the hurdle and belong to the Life Sciences sector show a higher level of production than those belonging to the Social Sciences and Humanities sector. This result also holds for the Physical and Engineering Sciences sector for the *h_index* variable. Analogously, in the zero component part of the models, the odds of the non-active status is significantly smaller for the researchers of the Life Sciences sector. In particular, the relative increase of the odds of having a zero value for the researchers belonging to the Social Sciences and Humanities sector with respect to those belonging to Life Sciences sector is equal to 1.8% for *repository*, to 127% for *scopus*, and to 185% for

*h_index*; while the relative increase is about 2% for all the three variables when the Physical and Engineering Sciences sector is considered. However, such results should be cautiously interpreted owing to the different coverage of the Scopus database, as well as to the different publication and citation patterns in the sectors (Iglesias and Pecharroman 2007). As is well known, the Scopus database has a weaker coverage for Social Sciences and Humanities, especially for non-English language countries, such as Italy. In the Social Sciences and Humanities sector, research results are communicated to the academic community mainly by means of books and chapters in books. Therefore, for this sector, the bibliographic databases, such as Scopus, are largely incomplete in terms of publications and citations (Hicks 2004). It is also worth remembering that co-authorship patterns are very different across scientific sectors. It can be argued that for sectors where articles have usually dozens of authors, the probability to be non-active is lower than in other sectors - such as Social Sciences and Humanities - where groups of co-authors are very small and single authorship often prevails. We also performed the regression analysis separately on the three datasets obtained by considering the different ERC sectors (Social Sciences and Humanities, Life Sciences, and Physical and Engineering Sciences). The corresponding results per discipline (available from the authors upon request) do not reveal marked different patterns with respect to those performed on the complete dataset.

As to the connection between teaching and research activities, when the zero component of the model is considered, a major teaching load in terms of teaching hours, number of students and thesis supervised does not affect the odds to be a non-active researcher. Indeed, our findings tend to reject the hypothesis that the odds to be non-active is affected by the crowding out effect between teaching and publications. Analogous conclusions have been achieved - using a completely different approach - by Kelchtermans and Veugelers (2011).

However, if active researchers are considered, it is worth noting that, concerning the three explanatory variables adopted to proxy teaching tasks (i.e. *teaching*, *thesis* and *students*), only *students* weakly significantly affects the research performance in terms of a weak decrease in the number of publication in the Scopus and *h*-index value, (relative decreases respectively equal to 0.3% and 0.2%. On the contrary, when the number of publications in the repository is considered, there is a weak evidence that the variable *students* positively, although moderately, influences the research output (relative increase equal to 0.2%); and a strong evidence that the number of thesis increases the number of publications, with a relative increase equal to 3.4%. It may be argued that the results achieved in the thesis can be used by supervisors to produce a research outcome suitable to appear in the repository - which, it is worth to remember, includes any type of publication - but not in peer-reviewed journals indexed by the Scopus database. Our results, differently from previous analyses (Fox 1992; Taylor et al. 2006; Kossi et al. 2013), suggest that neither substitution nor complementarity simple hypotheses seem to adequately represent the multifaceted relation among research and teaching activities.

As to the connection between the administrative duties of a researcher and her/his research activities, we found that the participation to Faculty meetings has a significant negative effect on the number of publications in the Scopus database for active researchers. Moreover, it increases the odds of having zero h-index value, while significantly reduces the odds of having zero publications on the research repository. These results seem to suggest that researchers productivity is negatively affected by bureaucratic and administrative tasks - a topic not covered in previous literature and deserving more scrutiny.

A positive evaluation received for the Prin projects significantly increases (i) the expected output of active researchers, as highlighted by the positive coefficient estimates in the count component of the model; and (ii) the probability of passing the hurdle, as shown by the negative sign of the estimates in the zero part of the model. Also in this case we are not able to interpret causally these results. In fact we cannot exclude that more productive researchers may be more likely to obtain positive evaluation.

As for the labour contract of researchers, a part-time contract significantly increases the probability to be non-active, with the most marked effect when the variable *h_index* is considered. However, it is worth noting that, among the active researchers, a part-time contract does not significantly affect the research performance. About this mixed evidence, it is possible to conjecture that some part-time-researchers are engaged in scientific research and achieved results similar to those of their full-time colleagues; others are devoted mainly to private practice outside university, and therefore they are non-active researchers.

As to the features of the department a researcher belongs to, it is worth noting that the relative number of researchers in more junior ranks and researchers in the same or more senior ranks does not seem to impact the odds to be non-active for *repository* and *scopus*, while it does for *h_index*. The same variable seems to have a significant positive effect for the *h_index* variable and a moderately significant negative effect for the *scopus* variable when active researchers are considered. These results suggest another puzzling question about the relation between the composition of departments and researchers' productivity (Carayol and Matt 2006; Stephan 2012). A possible but not exhaustive conjecture is that the presence of productive junior researchers widen the citation network of senior researchers improving the number of citations received by the latters and thus affecting their h-index value.

The composition of the department, in terms of the ratio of the number of non-research staff units to the number of permanent researchers, affects both the odds to be non-active and the value of the production indicators, thus evidencing a positive effect of staff availability on scientific productivity. In particular, an increase in the ratio between technical and administrative workers and researchers (*taw_ratio*) has a significant positive effect on the production of active researchers when *repository* or *h_index* variables are considered, with a relative increase approximately equal to 56% and to 45% respectively. It also has a significant effect on the reduction of the odds to be a non-active researcher, with a relative decrease of the odds equal to 74% for *repository* and 86% for *h-index*. It is worthwhile to remember that departmental staff consists of administrative personnel and specialized technicians directly involved in research activities coordinated by academic staff. This is particularly true in experimental sciences departments. Thus, on one hand we can suppose that administrative staff conveniently help a researcher in administrative and bureaucratic tasks connected with teaching and with the management of projects and research activities. On the other hand, we can suppose that the contribution of technical workers could be particularly relevant for those scholars working in experimental sciences departments where laboratory activities are the core of scientific research.

## 5. Conclusions

This article contributes to the stream of literature investigating the individual determinants of researcher performance. We analyse original data referring to a medium-sized Italian university which employs 942 researchers covering many scientific fields in Life Sciences, Physical and Engineering Sciences and Social Sciences and Humanities. All the researchers have a permanent position in the university and obey the same rules defined at a national level for recruitment, career, didactic charges, administrative duties and wages. Data refers to a three-year time-span (2008-2010).

With respect to previous literature, we adopt eight regression models to manage skewed count data with an excess of zero-valued observations, which are often the main features of the response variables adopted to quantify research production. Among these models, the Hurdle Negative Binomial exhibits a good fitting and appears to be reasonably coherent with the underlying generating data process. This model can be interpreted as follows: all the researchers are considered as potentially productive; when a researcher passes the hurdle by writing a paper and becoming active in the time period considered, her/his performance is described by the count component of the model. Moreover, the odds to be non-active is modelled by the zero component of the model. In order to highlight the determinants of a researcher production, we introduced an extensive set of explanatory variables, which, to the best of our knowledge, were not jointly explored by previous literature in an unique framework. A first result, widely known in literature, is that the different publication and citation practices adopted in different research fields have a strong impact on productivity of active researcher and on the odds to be non-active. This strong evidence suggests to consider a research priority the investigation of the processes giving rise to these sectorial differences, which are substantially unexplained in our model .

Results regarding personal observable characteristics of researchers confirm the evidences of previous studies. In particular, our data add evidence to the gender productivity puzzle and to the age-effect in publication activity. Academic position positively influences researcher's productivity, while seniority tends to have a negative effect. The evidence of our analysis does not allow to draw a clear-cut conclusion about the relation between teaching and research activities. A major teaching load does not affect the odds to be a non-active researcher and it does not reduce significantly the number publications for active researchers. The number of thesis supervised increases the number of publications in the repository. This evidence suggests that no univocal substitution or complementarity hypothesis can be claimed. Also in this case an in-deep analysis of factual evidence and underlying processes is straightforward to gain a better understanding of the phenomena.

 On the contrary, a clear result emerges about administrative tasks. These appear to negatively affect research productivity, especially when research outcomes filtered by a reviewing process are considered. To the best of our knowledge, this is a new piece of evidence that deserves further scrutiny. Our data also allowed to analyse the effect of the departmental working context on productivity. A first clear result is that operating in a department filled with more administrative and technical staff enhances productivity. On the contrary, mixed evidences emerge when the composition of the department in terms of senior/junior researchers is considered.

However, we caution against generalizations of our results since the adopted database include scientists working in a single university and we are aware that more general country-level investigation on the main determinants of scientific production should be undertaken. We believe that further research is needed in order to create improved measures of the effort devoted to institutional duties and university governance and to better understand if and how these activities could affect scientific production. Similarly, additional proxies for the teaching load have to be exploited. To this purpose, future research could surely benefit from the availability of information concerning researcher's final outputs (publications, patents, products, etc.), but also projects, scientific areas, teaching and administrative and institutional activities. Indeed, the evaluation of the effects of administrative and teaching tasks on scientific output is mandatory to verify if current incentive policies for stimulating research production are effective.

**Fig. 1** The observed distribution of the number of publications in the research repository of the University of Siena along with the predicted distribution obtained using the HNB model. Values greater than 40 are omitted for improving readability

**Fig. 2** The observed distribution of the number of publications in Scopus along with the predicted distribution obtained using the HNB model. Values greater than 40 are omitted for improving readability

**Fig. 3** The observed distribution of the h-index value along with the predicted distribution obtained using the HNB model. Values greater than 40 are omitted for improving readability

**Table 1**. Definition of variables. Mean, $1^{st}$ quartile, median, $3^{rd}$ quartile, minimum and maximum values, Coefficient of Variation (C.V.) are reported for quantitative variables; relative frequencies are reported for categorical variables.

| Response variables | Description | Mean | $1^{st}$ Qu | Median | $3^{rd}$ Qu | Min. | Max. | C.V. |
|---|---|---|---|---|---|---|---|---|
| *repository* | number of publications in Anagrafe.UNISI | 8.47 | 2.00 | 6.00 | 11.00 | 0 | 87.00 | 1.11 |
| *scopus* | number of publications in Scopus database | 6.45 | 0.00 | 2.00 | 8.00 | 0 | 176.00 | 2.11 |
| *h_index* | h- index | 6.12 | 0.00 | 3.00 | 10.00 | 0 | 57.00 | 1.28 |

| Quantitative explanatory variables | Description | Mean | $1^{st}$ Qu | Median | $3^{rd}$ Qu | Min. | Max. | C.V. |
|---|---|---|---|---|---|---|---|---|
| *age* | age of the scholar | 53.10 | 45.93 | 53.77 | 60.80 | 31.72 | 70.69 | 0.17 |
| *tenure* | years in permanent position | 18.11 | 9.99 | 18.00 | 29.00 | 2.00 | 42.00 | 0.58 |
| *teaching* | average number of hours of teaching per month worked | 7.39 | 5.00 | 7.22 | 9.39 | 0 | 52.00 | 0.53 |
| *thesis* | number of thesis as supervisor | 2.60 | 0.33 | 1.30 | 3.67 | 0 | 44.00 | 1.51 |
| *students* | number of students attending lectures | 35.16 | 11.54 | 26.50 | 48.92 | 0 | 415.00 | 0.98 |
| *presence_faculty_meeting* | proportion of presence to the Faculty meeting | 0.50 | 0.28 | 0.55 | 0.74 | 0 | 1.00 | 0.55 |
| *more_junior_ratio* | ratio of researchers in more junior ranks and researchers in the same or more senior ranks | 2.98 | 1.11 | 2.02 | 4.14 | 0 | 20 | 0.97 |
| *taw_ratio* | ratio of technical and administrative workers and permanent researchers | 0.51 | 0.17 | 0.43 | 0.75 | 0.08 | 1.41 | 0.71 |

| Categorical explanatory variables | Description | Frequency |
|---|---|---|
| *gender* | gender | |
| M | male | 0.66 |
| F | female | 0.34 |
| *position* | role at the university | |
| Assistant professor | assistant professor | 0.41 |
| Associate professor | associate professor | 0.29 |
| Full professor | full professor | 0.30 |
| *erc* | European Research Council Sectors | |
| LS | Life Sciences | 0.29 |
| PE | Physical and Engineering Sciences | 0.28 |
| SH | Social Sciences and Humanities | 0.43 |
| *prin* | positive evaluation in Prin project | |
| 0 | no positive evaluation | 0.26 |
| 1 | yes positive evaluation | 0.74 |
| *full-time* | full-time service to the university | |
| 0 | no full-time | 0.06 |
| 1 | yes full-time | 0.94 |

**Table 2**. Correlation matrix of the quantitative explanatory variables.

| | *age* | *tenure* | *teaching* | *thesis* | *students* | *presence_ faculty_meeting* | *more_junior_ ratio* | *taw_ratio* |
|---|---|---|---|---|---|---|---|---|
| *age* | 1.00 | 0.83 | 0.06 | -0.06 | -0.07 | -0.07 | 0.36 | 0.07 |
| *tenure* | 0.83 | 1.00 | 0.04 | -0.07 | -0.04 | -0.11 | 0.29 | 0.13 |
| *teaching* | 0.06 | 0.04 | 1.00 | 0.21 | 0.26 | 0.11 | 0.14 | -0.08 |
| *thesis* | -0.06 | -0.07 | 0.21 | 1.00 | 0.29 | 0.13 | 0.05 | -0.19 |
| *students* | -0.07 | -0.04 | 0.26 | 0.29 | 1.00 | 0.25 | 0.13 | -0.09 |
| *presence_ faculty_meeting* | -0.07 | -0.11 | 0.11 | 0.13 | 0.25 | 1.00 | 0.12 | -0.13 |
| *more_junior_ratio* | 0.36 | 0.29 | 0.14 | 0.05 | 0.13 | 0.12 | 1.00 | 0.14 |
| *taw_ratio* | 0.07 | 0.13 | -0.08 | -0.19 | -0.09 | -0.13 | 0.14 | 1.00 |

**Table 3**. Log-likelihood (log-L) and Akaike's information criterion (AIC) values of the fitted models for the three response variables.

| Models | *repository* | | *scopus* | | *h_index* | |
|--------|--------|--------|--------|--------|--------|--------|
|        | log-L | AIC | log-L | AIC | log-L | AIC |
| P | -4383.68 | 8799.40 | -4105.30 | 8242.60 | -2508.68 | 5049.40 |
| NB | -2859.80 | 5753.60 | -2103.72 | 4241.50 | -2091.33 | 4216.70 |
| ZIP | -4065.00 | 8194.01 | -3668.00 | 7400.49 | -2283.00 | 4629.15 |
| ZINB | -2832.00 | 5729.32 | -2040.00 | 4146.00 | -1981.00 | 4028.75 |
| HP | -4065.00 | 8194.43 | -3668.00 | 7399.80 | -2281.00 | 4626.91 |
| HNB | -2830.00 | 5726.68 | -2038.00 | 4141.08 | -1983.00 | 4032.47 |
| W | -2846.57 | 5729.14 | -2041.44 | 4118.88 | -1994.02 | 4024.04 |
| S | -2858.43 | 5752.86 | -2081.16 | 4198.32 | -2070.18 | 4176.36 |

**Table 4**. Observed and expected frequencies for the ZINB, HNB, W models for the three response variables.

| Variable | Value | Observed frequency | Expected frequency | | | Value | Observed frequency | Expected frequency | | |
|---|---|---|---|---|---|---|---|---|---|---|
| | | | ZINB | HNB | W | | | ZINB | HNB | W |
| *repository* | 0 | 103 | 106.88 | 103.00 | 73.04 | 13-15 | 61 | 62.02 | 61.55 | 54.84 |
| | 1 | 80 | 69.03 | 72.04 | 91.12 | 16-18 | 40 | 42.45 | 42.15 | 37.39 |
| | 2 | 57 | 73.99 | 75.77 | 90.62 | 19-21 | 22 | 29.33 | 29.15 | 26.14 |
| | 3 | 67 | 72.46 | 73.39 | 83.35 | 22-25 | 25 | 25.89 | 25.77 | 23.70 |
| | 4-6 | 203 | 185.67 | 186.07 | 195.05 | 26-29 | 12 | 16.39 | 16.35 | 15.69 |
| | 7-9 | 133 | 132.77 | 132.11 | 126.86 | 30-34 | 13 | 12.64 | 12.63 | 12.84 |
| | 10-12 | 101 | 91.06 | 90.40 | 82.50 | ≥35 | 25 | 21.42 | 21.63 | 28.86 |
| *scopus* | 0 | 394 | 394.91 | 394.00 | 381.78 | 13-15 | 34 | 34.40 | 34.50 | 35.44 |
| | 1 | 54 | 66.87 | 67.90 | 82.05 | 16-18 | 16 | 25.60 | 25.64 | 25.70 |
| | 2 | 52 | 56.70 | 56.82 | 58.05 | 19-21 | 14 | 19.41 | 19.42 | 18.93 |
| | 3 | 51 | 47.62 | 47.49 | 47.06 | 22-25 | 20 | 19.14 | 19.11 | 18.05 |
| | 4-6 | 124 | 104.15 | 103.87 | 104.74 | 26-29 | 10 | 13.77 | 13.73 | 12.53 |
| | 7-9 | 77 | 68.15 | 68.22 | 70.94 | 30-34 | 10 | 12.16 | 12.11 | 10.69 |
| | 10-12 | 56 | 47.46 | 47.60 | 49.67 | ≥35 | 30 | 31.66 | 31.59 | 26.38 |
| *h_index* | 0 | 351 | 347.36 | 351.00 | 335.60 | 13-15 | 62 | 52.22 | 53.27 | 50.83 |
| | 1 | 55 | 42.44 | 41.22 | 72.57 | 16-18 | 42 | 35.46 | 36.07 | 36.01 |
| | 2 | 44 | 46.77 | 45.30 | 50.85 | 19-21 | 27 | 23.65 | 23.92 | 25.02 |
| | 3 | 44 | 46.34 | 44.89 | 43.65 | 22-25 | 17 | 19.60 | 19.65 | 21.64 |
| | 4-6 | 103 | 125.18 | 122.79 | 112.36 | 26-29 | 11 | 11.33 | 11.23 | 13.03 |
| | 7-9 | 86 | 100.04 | 100.45 | 90.73 | 30-34 | 5 | 7.81 | 7.65 | 9.26 |
| | 10-12 | 86 | 74.33 | 75.60 | 69.53 | ≥35 | 9 | 9.46 | 8.96 | 10.91 |

**Table 5**. Estimated parameters, Standard Errors (S.E.) and Significance (Sig.), log-likelihood (log-L) and Akaike's information criterion (AIC) values for the HNB model.

| Variables | repository | | | scopus | | | h_index | | |
|---|---|---|---|---|---|---|---|---|---|
| | Estimate | S.E. | Sig. | Estimate | S.E. | Sig. | Estimate | S.E. | Sig. |
| *Count component of the model* | | | | | | | | | |
| Intercept | 3.388 | 0.396 | *** | 3.876 | 0.596 | *** | 3.250 | 0.381 | *** |
| *age* | -0.023 | 0.007 | *** | -0.024 | 0.010 | * | -0.018 | 0.007 | ** |
| *gender*_F | Ref. | | | | | | | | |
| *gender*_M | 0.160 | 0.067 | * | 0.230 | 0.099 | * | 0.014 | 0.062 | |
| *position*_Full professor | Ref. | | | | | | | | |
| *position*_Associate professor | -0.381 | 0.102 | *** | -0.391 | 0.159 | * | -0.265 | 0.097 | ** |
| *position*_Assistant professor | -0.879 | 0.135 | *** | -0.954 | 0.209 | *** | -0.660 | 0.128 | *** |
| *erc*_LS | Ref. | | | | | | | | |
| *erc*_PE | -0.008 | 0.085 | | -0.111 | 0.102 | | -0.193 | 0.062 | ** |
| *erc*_SH | -0.243 | 0.097 | * | -1.924 | 0.184 | *** | -1.637 | 0.127 | *** |
| *tenure* | -0.009 | 0.005 | | -0.020 | 0.008 | * | -0.006 | 0.005 | |
| *teaching* | -0.016 | 0.010 | | -0.004 | 0.012 | | 0.002 | 0.007 | |
| *thesis* | 0.033 | 0.008 | *** | 0.012 | 0.016 | | -0.013 | 0.011 | |
| *students* | 0.002 | 0.001 | * | -0.003 | 0.001 | * | -0.002 | 0.001 | * |
| *prin*_0 | Ref. | | | | | | | | |
| *prin*_1 | 0.269 | 0.079 | *** | 0.740 | 0.119 | *** | 0.411 | 0.073 | *** |
| *full-time*_0 | Ref. | | | | | | | | |
| *full-time*_1 | -0.068 | 0.148 | | 0.077 | 0.209 | | -0.014 | 0.128 | |
| *presence_faculty_meeting* | 0.058 | 0.125 | | -0.675 | 0.175 | *** | -0.066 | 0.108 | |
| *more_junior_ratio* | -0.020 | 0.016 | | -0.049 | 0.023 | * | 0.039 | 0.014 | ** |
| *taw_ratio* | 0.445 | 0.112 | *** | 0.156 | 0.144 | | 0.374 | 0.086 | *** |
| Log(theta) | 0.541 | 0.076 | *** | 0.278 | 0.094 | ** | 1.355 | 0.098 | *** |
| *Zero component of the model* | | | | | | | | | |
| Intercept | -2.748 | 1.535 | . | -6.973 | 1.323 | *** | -9.050 | 1.487 | *** |
| *age* | 0.028 | 0.028 | | 0.074 | 0.022 | *** | 0.106 | 0.025 | *** |
| *gender*_F | Ref. | | | | | | | | |
| *gender*_M | -0.048 | 0.262 | | -0.315 | 0.223 | | -0.477 | 0.257 | . |
| *position*_Full professor | Ref. | | | | | | | | |
| *position*_Associate professor | 0.086 | 0.425 | | 1.315 | 0.335 | *** | 1.541 | 0.363 | *** |
| *position*_Assistant professor | 0.857 | 0.515 | . | 2.447 | 0.444 | *** | 3.036 | 0.491 | *** |
| *erc*_LS | Ref. | | | | | | | | |
| *erc*_PE | 1.126 | 0.395 | ** | 1.142 | 0.346 | *** | 1.097 | 0.494 | * |
| *erc*_SH | 0.732 | 0.406 | . | 4.854 | 0.415 | *** | 5.227 | 0.512 | *** |
| *tenure* | 0.047 | 0.022 | * | 0.028 | 0.018 | | -0.009 | 0.020 | |
| *teaching* | 0.033 | 0.028 | | 0.017 | 0.028 | | 0.022 | 0.041 | |
| *thesis* | -0.027 | 0.039 | | 0.007 | 0.025 | | 0.015 | 0.026 | |
| *students* | -0.002 | 0.005 | | -0.002 | 0.004 | | -0.001 | 0.004 | |
| *prin*_0 | Ref. | | | | | | | | |
| *prin*_1 | -1.294 | 0.263 | *** | -1.070 | 0.267 | *** | -1.009 | 0.307 | ** |
| *full-time*_0 | Ref. | | | | | | | | |
| *full-time*_1 | -0.685 | 0.428 | | -0.773 | 0.396 | . | -1.347 | 0.473 | ** |
| *presence_faculty_meeting* | -1.280 | 0.433 | ** | -0.157 | 0.386 | | 1.187 | 0.431 | ** |
| *more_junior_ratio* | -0.170 | 0.100 | . | 0.068 | 0.058 | | 0.156 | 0.064 | * |
| *taw_ratio* | -1.334 | 0.488 | | -0.734 | 0.394 | . | -1.947 | 0.556 | *** |
| log-L | -2830.000 | | | -2038.000 | | | -1983.000 | | |
| AIC | 5726.676 | | | 4141.081 | | | 4032.468 | | |
| Number of observations | 942 | | | 942 | | | 942 | | |

. Significant at 10%
* Significant at 5%
** Significant at 1%
*** Significant at 0.1%

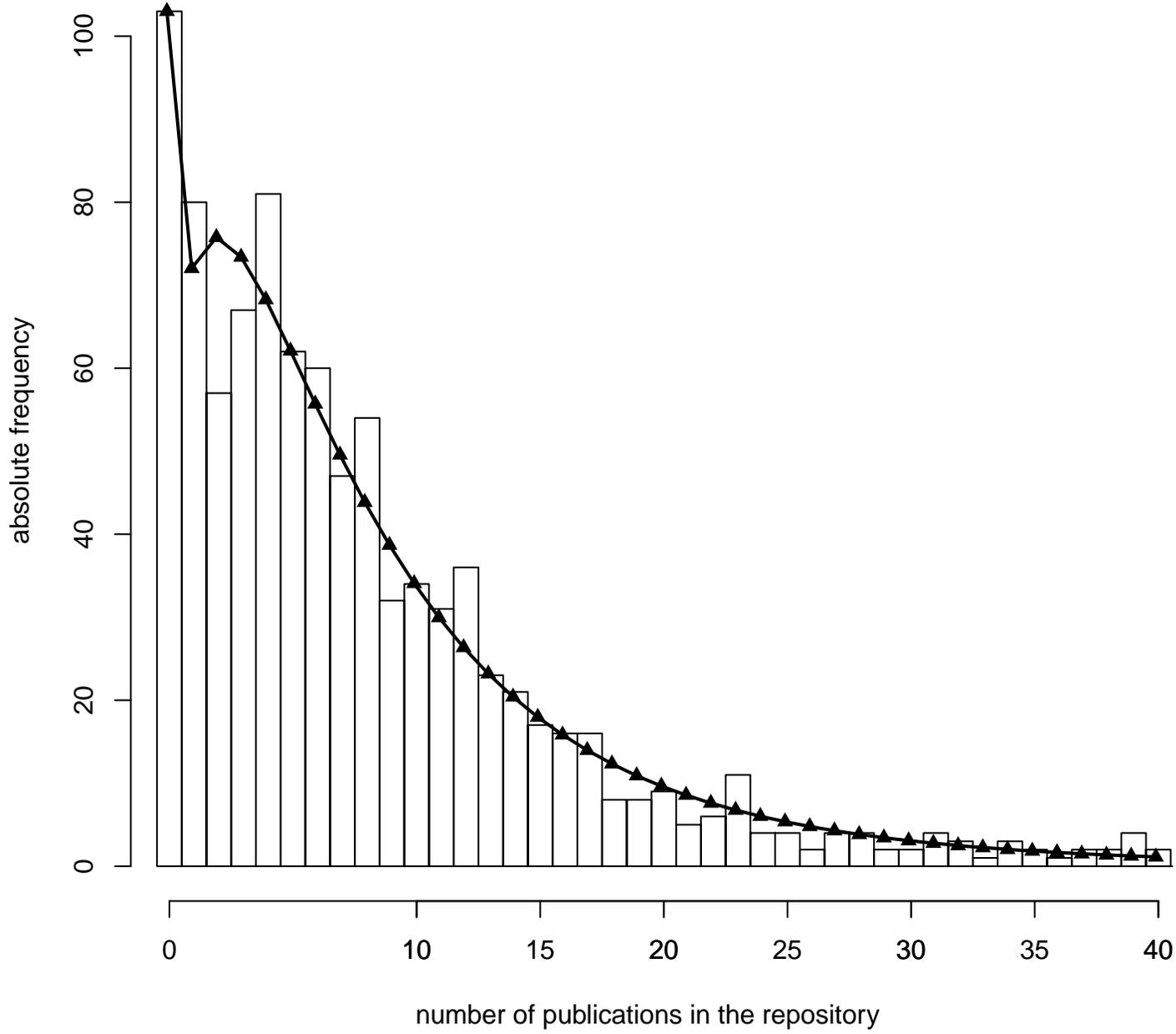

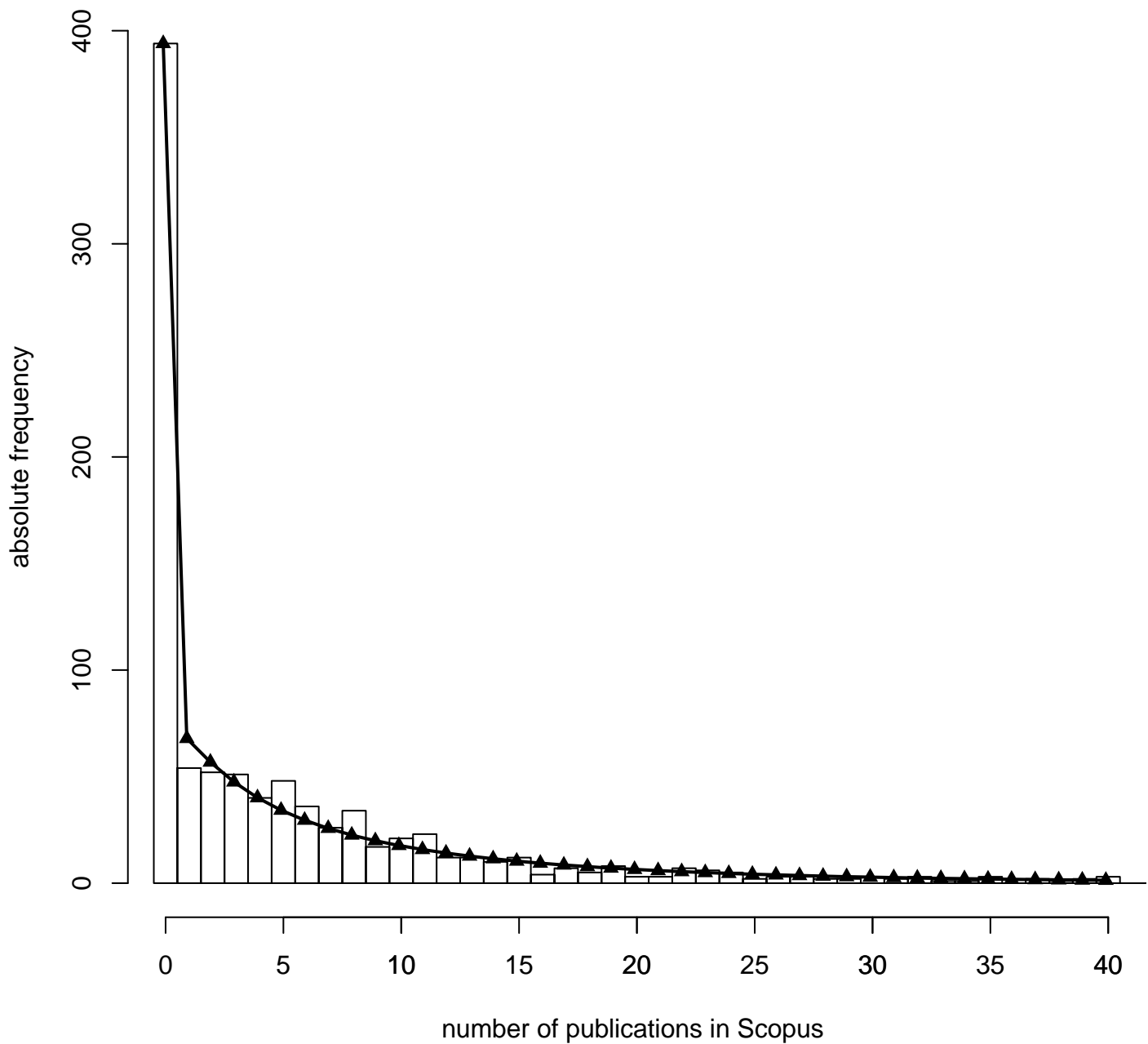

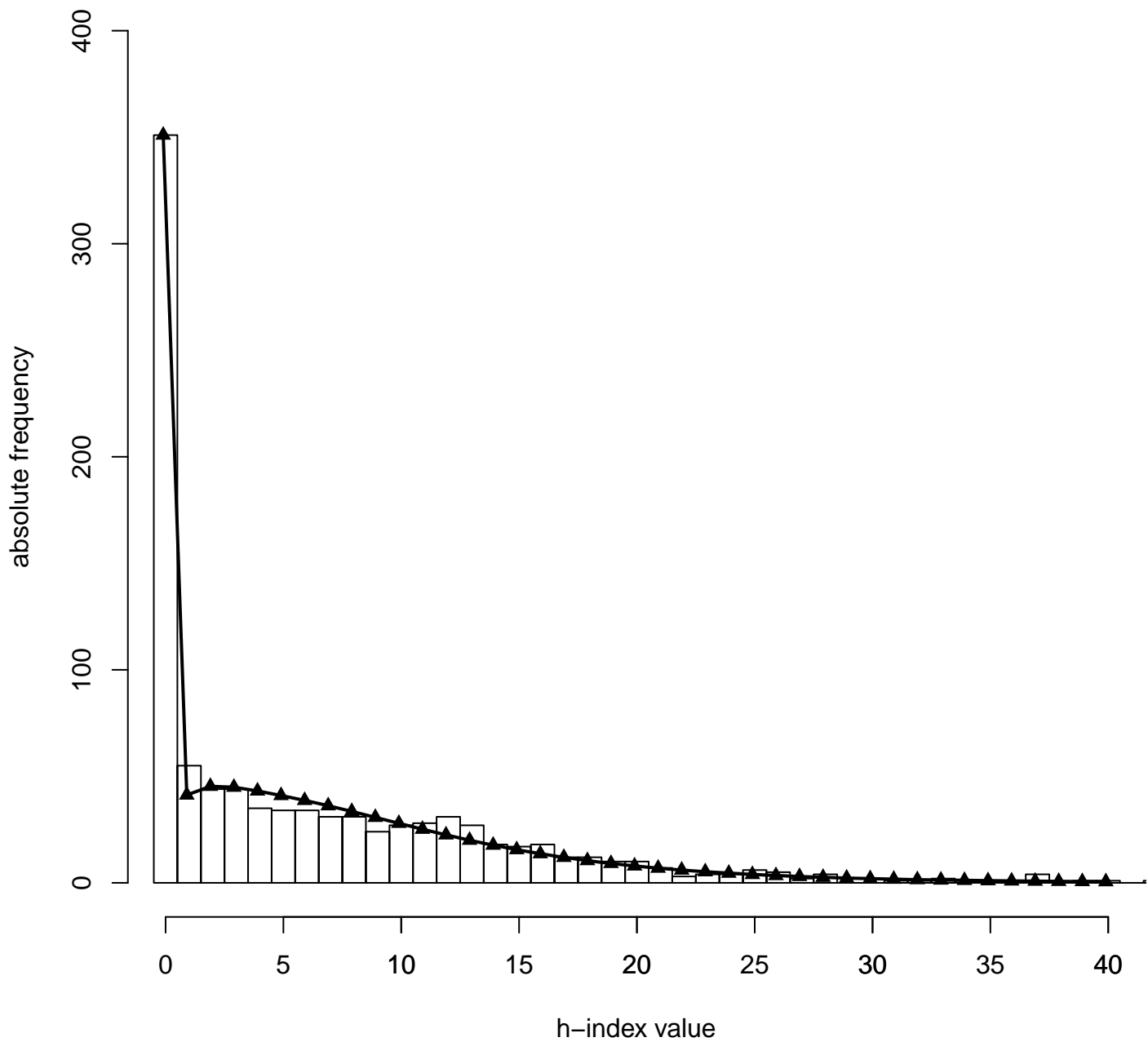